\pdfoutput=1

\documentclass[11pt]{article}

\usepackage{acl}

\usepackage{times}
\usepackage{latexsym}
\usepackage{graphicx}
\usepackage{multirow}
\usepackage{amsmath}
\usepackage{float}
\usepackage{hyperref}

\usepackage[T1]{fontenc}

\usepackage[utf8]{inputenc}

\usepackage{microtype}

\usepackage{inconsolata}

\title{Language-based Valence and Arousal Expressions between the United States and China: a Cross-Cultural Examination}

\author{
Young-Min Cho$^1$ \quad Dandan Pang$^2$ \quad Stuti Thapa$^3$ \quad Garrick Sherman$^1$ \\ \quad \textbf{Lyle Ungar}$^1$ \quad \textbf{Louis Tay}$^4$ \quad \textbf{Sharath Chandra Guntuku}$^1$ \\
$^1$University of Pennsylvania \quad $^2$Bern University of Applied Sciences \\ \quad $^3$University of Tulsa  \quad $^4$Purdue University\\
\texttt{\{jch0,ungar,sharathg\}@seas.upenn.edu, dandan.pang@bfh.ch, stt0912@utulsa.edu}\\  
\texttt{garricktsherman@gmail.com, stay@purdue.edu}
}

\begin{document}
\maketitle
\begin{abstract}
While affective expressions on social media have been extensively studied, most research has focused on the Western context. This paper explores cultural differences in affective expressions by comparing valence and arousal on Twitter/X (geolocated to the US) and Sina Weibo (in Mainland China). Using the NRC-VAD lexicon to measure valence and arousal, we identify distinct patterns of emotional expression across both platforms. Our analysis reveals a functional representation between valence and arousal, showing a negative offset in contrast to traditional lab-based findings which suggest a positive offset. Furthermore, we uncover significant cross-cultural differences in arousal, with US users displaying higher emotional intensity than Chinese users, regardless of the valence of the content. Finally, we conduct a comprehensive language analysis correlating n-grams and LDA topics with affective dimensions to deepen our understanding of how language and culture shape emotional expression. These findings contribute to a more nuanced understanding of affective communication across cultural and linguistic contexts on social media.\footnote{For future research, we release our dataset at \href{https://github.com/JeffreyCh0/X_Weibo_Valence_Arousal}{https://github.com/JeffreyCh0/X\_Weibo\_Valence\_Arousal}}

\end{abstract}

\section{Introduction}

Subjective expressions of affect (how we feel) play a crucial role in understanding learning outcomes in individuals \cite{Hourihan2017-gs}, their perceptions \cite{Gorn2001-lf}, well-being \cite{Xu2015-mm}, and mental and physical health \cite{Cohen2006-vy}. Multiple theoretical and empirical works have, therefore, examined the underlying dimensions of affect and their relationships. While there are several models of affective structure, Russell's two-dimensional circumplex model is the most widely recognized, where orthogonal valence (pleasant to unpleasant) and arousal (high to low activation) are represented as the horizontal and vertical axes\footnote{See Russell's two-dimensional circumplex model in Figure \ref{fig:russell}.} \cite{Russell1980-au, Yik1999-ra}.

\begin{figure}[t]
    \centering
    \includegraphics[width=0.9\linewidth]{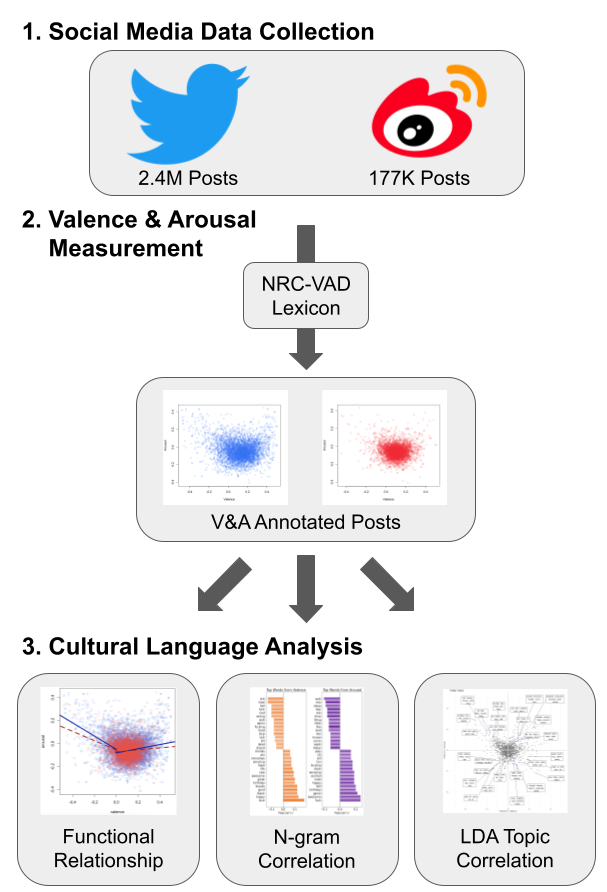}
    \caption{The analysis pipeline of this paper compares cultural differences in affective expressions using large-scale social media data. We examine the functional relationship between valence and arousal and explore the differences through language analysis methods.}
    \label{fig:spirit}
\end{figure}

Understanding the functional relationship between valence and arousal in this two-dimensional space is of empirical, psychometric, and theoretical interest. Among various models from previous studies \cite{Ortony1990-qi,Lang1994-xd}, a "V-shaped" relationship, where arousal is a function of valence, is one of the most widely tested and accepted \cite{Kuppens2013-im,Cacioppo1999-fk}. Arousal is shown to be directly related to the intensity of positive or negative valence with a positivity offset and a negativity bias, with varying levels of cross-cultural support \cite{kuppens2017relation}.

While the affective structure and the valence-arousal relationship are often considered universal, most previous studies focus exclusively on Western samples, overlooking cross-cultural heterogeneity \cite{Tsai2006-vc}. Different cultures value emotions (ideal affect) uniquely and adhere to distinct standards for emotional expression \cite{matsumoto1990cultural}. For instance, Americans tend to associate enthusiasm (high arousal) with positive valence, while Asians often prefer quietness (low arousal) \cite{Tsai2006-vc}. Although some studies have highlighted affective differences between Western and Eastern cultures, these are typically based on small, lab-based samples \cite{Kuppens2017-aa}, which may be biased by self-reporting and recall issues \cite{Tarrant1993-ds,Winograd2006-qb}. 

To address these limitations, researchers advocate for studies that go beyond self-reports and focus on behaviors \cite{baumeister2007psychology}, especially on social media. Social media data offer a naturalistic and ecological setting to capture individuals' emotions and, despite potential social desirability bias, have been shown to reliably estimate well-being \cite{Jaidka2020-ai, liou2023online}, sentiment \cite{Preotiuc-Pietro2016-kl}, and personality \cite{Schwartz2013-bj, havaldar-etal-2024-building}. 

This paper examines the cross-cultural difference on the affective expressions with functional relationship between valence and arousal by analyzing natural language expressions from Twitter/X (geolocated to the US) and Sina Weibo (in Mainland China) posts. The pipeline of our work is shown in Figure \ref{fig:spirit}. The study has three key contributions: 
\begin{itemize}
    \item We evaluate functional representations between valence and arousal on large-scale social media data, identifying a negative offset in contrast to previous lab-based studies.
    \item We demonstrate cross-cultural differences in valence and arousal, showing that US users exhibit stronger emotional intensity (higher arousal) than Chinese users across both positive and negative valence.
    \item We employ comprehensive language analysis, correlating n-grams and LDA topics with valence and arousal, providing insights into the functional relationship and cultural divergence in affective expression.
\end{itemize}

\section{Methods}

\subsection{Data Preparation}
Our data consist of public messages posted on Weibo and Twitter. The content and behavior variations on Weibo and Twitter have been studied in different contexts \cite{ma2013electronic,Lin2016-vw}. While working with non-random, non-representative samples poses challenges, social media posts can still reveal psychological traits, demographics \cite{Sap2014-tq,Zhang2016-ev}, location \cite{Salehi2017-qu,Zhong2015-rx}, and mental health \cite{guntuku2019twitter, Tian2018-jq}. 

To collect Twitter data, we used the survey platform Qualtrics, which included demographic questions such as gender and age. Participants from the US shared their Twitter handles after completing the survey. Users were compensated for their time, and consented to share their Twitter posts. There were 3,113 Twitter users, with around 3.6 million posts until 2016. 

Weibo, unlike Twitter, does not offer an API tool for obtaining random samples over time. So, starting with a random set of individuals from a public dataset \cite{Guntuku2019-vd}, Weibo posts were gathered using a breadth-first search method on users\footnote{We started with a random set of users, and expanded to all their friends (bidirectional, similar to followers + following), and we repeated the process.}. We obtained over 29 million posts from 2014 from 859,054 people on Weibo. Gender and age were collected from self-reported demographic information on their Weibo profile. Subsetting to users posted more than 500 words and with a reasonable self-reported age (<100 years) and gender, the dataset consisted of 668,257 Weibo posts from 8,731 users. 500 words were found to be the minimum threshold to obtain reliable psychological estimates from individuals' language \cite{Eichstaedt2021-py, Jaidka2018-xr}. 

Based on the gender and age distribution of Weibo and Twitter users, we built propensity-score-based matched samples, resulting in 2,191 users each on both platforms with at least 500 words\footnote{We use propensity score matching for its nuanced handling of continuous variables and its allowance for quantitative assessment of covariate balance between matched groups.} \cite{rosenbaum1983central}. These matched users had 2.4 million posts on Twitter and 177,042 posts on Weibo. In our matched dataset 67.1\% self-reported as being female and 32.9\% as male, and the mean age was 26.9 (s.d. 8.8). On Twitter, there were on average 15.6 (s.d. 2.8) words per user and Weibo had 57.3 (s.d. 15.4) words per user. The differences in word counts are driven by the lack of character limits to posts on Weibo. 

To eliminate the confounds of bilingualism \cite{Fishman1980-xc}, we retain only English posts on Twitter and Mandarin posts on Weibo by using langid \cite{Lui2012-um}. Re-tweets are also removed from both datasets ('RT @USERNAME:' on Twitter and '@USERNAME//' on Weibo). Weibo posts were split into tokens using THULAC \cite{Li2009-sv} while Twitter posts were segmented using happierfuntokenizing (DLATK/happierfuntokenizing, 2017) due to their ability to handle emoticons and other social media slang. To eliminate uncommonly used words (outliers), we filtered words with different frequency thresholds for each platform. Words used by fewer than 0.1\% of the total posts
on Twitter and 0.5\% on Weibo were removed from the analysis. Most words are seldom used in language, as they follow a Zipfian distribution. By removing these words, we ensure that the language insights from our research can be generalized to out-of-sample cases. 

\subsection{Valence \& Arousal Measurement}
The circumplex and vector models of emotion have been broadly used for representing affective states\footnote{See Russell's two-dimensional circumplex model in Figure \ref{fig:russell}.} \cite{Russell1980-au,Bradley1992-cu}. In these two-dimensional models, valence is the x-axis, expressing pleasantness and unpleasantness, attractiveness and aversiveness, joy, and sorrow \cite{Frijda1986-uy}. Arousal is the y-axis, describing the degree of wakefulness, boredom, excitement, and calm. These models allow any affective state, emotion, word, or expression to be represented as a point in the space, regardless of the difference in language, country, or culture. 

We measure valence and arousal using a validated data-driven lexicon generated based on the circumplex model in both English and Mandarin. We used NRC Valence, Arousal, and Dominance (NRC-VAD) Lexicon \cite{Mohammad2018-ko} for Twitter data and its translated version for Weibo data. NRC-VAD consists of valence and arousal weights for more than 20,000 words in English and shows a "V-shaped" relationship between two dimensions: extremely positive or negative valence is usually paired with high arousal, while calmness matches low arousal. We subtract 0.5 from all scores to make them zero-centered.

Multilinguality is another reason to choose NRC-VAD as our valence-arousal measurement lexicon. There are over 100 languages available for NRC-VAD (August 2022), and the authors claim that most affective norms are stable across languages. Since an original-translated term pair has the same scores, this lexicon avoids the annotator agreement and scale-matching issue, which are common problems using two different lexica over two languages. 

We calculate the valence and arousal scores for each post on Twitter and Weibo using NRC-VAD lexica. For each post, we sum the result of item-wise multiplication of relative word frequency within the post, and the corresponding valence or arousal score for the word. In detail, we follow the formula: 

\begin{equation}
    valence_m = \sum_{w \in W_m}valence_w \cdot \frac{freq_m(w)}{W_m}
\end{equation}
Where $m$ represents a post, $w$ is a word, $freq_m(w)$ is the frequency of word $w$ in post $m$, $valence_m$ and $valence_w$ are the valence scores for post $m$ (for annotation) and word $w$ (from the NRC-VAD lexicon), respectively, and $W_m$ is the total number of words in post $m$. Arousal is calculated in a similar manner.

\subsection{Evaluation of Functional Relationship}

\citealp{Kuppens2013-im} showed six possible functional relationships between valence and arousal. These models are independence (Model 1), Linear Relation (Model 2), Symmetric V-Shaped Relation (Model 3), and Asymmetric V-Shaped Relations, including asymmetric interception (Model 4), asymmetric slope (Model 5), and asymmetric interception and slope (Model 6). The models' functional representations are shown below\footnote{See full examples on our dataset on Figure \ref{fig:func_rela_full}.}:
\\

\resizebox{0.95\linewidth}{!}{$
A_m = 
\begin{cases}
    \beta_0 + \epsilon_m & \text{(Model 1)} \\
    \beta_0 + \beta_1V_m + \epsilon_m & \text{(Model 2)} \\
    \beta_0 + \beta_1|V_m| + \epsilon_m & \text{(Model 3)} \\
    \beta_0 + \beta_1|V_m| + \beta_2I_m + \epsilon_m & \text{(Model 4)} \\
    \beta_0 + \beta_1|V_m| + \beta_3I_m|V_m| + \epsilon_m & \text{(Model 5)} \\
    \beta_0 + \beta_1|V_m| + \beta_2I_m + \beta_3I_m|V_m| + \epsilon_m & \text{(Model 6)}
\end{cases}
$}
\\

Where $A_m$ and $V_m$ are short for $Arousal_m$ and $Valence_m$, arousal and valence scores for the post $m$, $I_m$ denotes a dummy variable that indicates whether $Valence_m$ is positive($I_{m}=1$) or negative($I_{m}=0$). Each model is tested with a within-person intercept and slope.

We use Akaike Information Criterion (AIC; \citealp{bozdogan1987model}), Bayesian Information Criterion (BIC; \citealp{Schwarz1978-vi}), posterior probability, and Conditional $R^2$ for model selection. AIC is defined as $AIC = 2 \cdot k -2 \cdot \ln(\hat{L})$, and BIC has the following format: $BIC = -2 \cdot \ln(\hat{L}) + k \cdot \ln(N)$, where $\hat L$ is the maximized value of the likelihood function of the model, k is the number of parameters, and N is the number of observations. One advantage of using BIC is that it can be used to approximate posterior probability for each model: 
\begin{equation}
    P(model_i|data) =\frac{exp(-0.5BIC_i)}{\sum exp(-0.5BIC_i)}
\end{equation}

While applying the six models, we use mixed effects models to fit the datasets. We assume there is a fixed relationship between valence and arousal across all posts, while the average level of arousal may vary from user to user. The regression models can correctly represent the relationship between the two variables by setting within-person differences as the random effect. In the model comparison, to cover both fixed and random effects, we use conditional $R^2$ to represent the proportion of variance explained by the entire model.

\subsection{Social Media Language Analysis}

\paragraph{Feature Extraction}
We extract two open-vocabulary features from Twitter and Weibo: n-grams and topics. N-grams help capture common word patterns and phrase structures that reflect how emotions are expressed in everyday language, allowing us to identify culturally specific linguistic cues tied to affective dimensions. Meanwhile, Latent Dirichlet Allocation (LDA, \citealp{blei2003latent}) uncovers underlying topics within the text, revealing thematic contexts that influence emotional expression. We choose LDA for topic modeling because it has better explainability and computational efficiency than other modern models like Top2Vec and BERTopic \cite{angelov2020top2vec, grootendorst2022bertopic}. By analyzing both n-grams and topics, we can better understand the interplay between language, culture, and emotion, providing a richer, data-driven perspective on cross-cultural affective communication.

 We collect contiguous sequences of one or two words (1-2 grams, \citealp{Kern2014-ik,Andrew_Schwartz2013-bb}) with pointwise mutual information (PMI = 3; \citealp{Church1990-ga}). This resulted in unique unigrams and bigrams set of 10,477 for Weibo and 12,798 for Twitter. We extracted the normalized distribution of the n-grams for each post in the Weibo and Twitter datasets. We then used 2,000 topics generated using LDA as the second feature to represent users' language in our Twitter and Weibo datasets \cite{schwartz2013personality}. We utilized topics generated on much larger datasets to favor high diversity and coverage. 2,000 English topics generated a corpus of approximately 18 million Facebook updates with alpha set to 0.30 to favor fewer topics per document. These have been shown to perform well across multiple platforms \cite{eichstaedt2015psychological}. 2,000 Mandarin topics were generated on 29 million Weibo posts with similar parameters set in Mallet \cite{mccallum2002mallet}. Inherently, each topic is realized as a set of words with probabilities. Every post is thus scored in terms of its probability of containing each of the 2,000 topics, $p(topic, post)$, which is derived from their probability of containing a word, $p(word|post)$, and the probability of the words being in given topics, $p(topic|word)$. 

\paragraph{Differential Language Analysis}
To understand the functional relationship and cultural difference in valence and arousal, we utilized ordinary least squares (OLS) regression to model valence and arousal on post level, controlling for gender and age by matching these variables between the Twitter and Weibo samples. The inputs for the regression were different language features independently extracted from social media posts - n-grams and topics; and outputs were valence and arousal, each of which was constructed using a separate OLS model. From OLS regression, we extracted coefficients to represent Pearson correlation coefficients for each feature dimension\footnote{For details of how Pearson $r$ is calculated, see Appendix \ref{app:pearson}}. To correct for multiple comparisons and control the false discovery rate in multiple hypothesis testing, we applied the Benjamini-Hochberg p-correction \cite{Benjamini1995-qg}. We considered correlations meaningful if they met the threshold of $p < .05$.

The formula for n-gram models are:
\begin{equation}
    valence_m \sim \sum_{n \in 1-2gram}a_n \cdot \frac{freq_m(n)}{N_m}+\varepsilon 
\end{equation}
where $m$ is a post, $n$ is an n-gram, $freq_m(n)$ is the frequency of $n$ in $m$, $a_n$ is coefficients and $N_m$ is a total number of n-grams in the post. 

The formula for LDA topic models are:
\begin{equation}
    valence_m \sim \sum_{t \in Topics}a_t \cdot P(t|m)+\varepsilon 
\end{equation}
where $P(t|m)$ is the probability of m belonging to topic t.
Models for arousal can be expressed in the same fashion.
\begin{figure}[t]
    \centering
    \includegraphics[width=\linewidth]{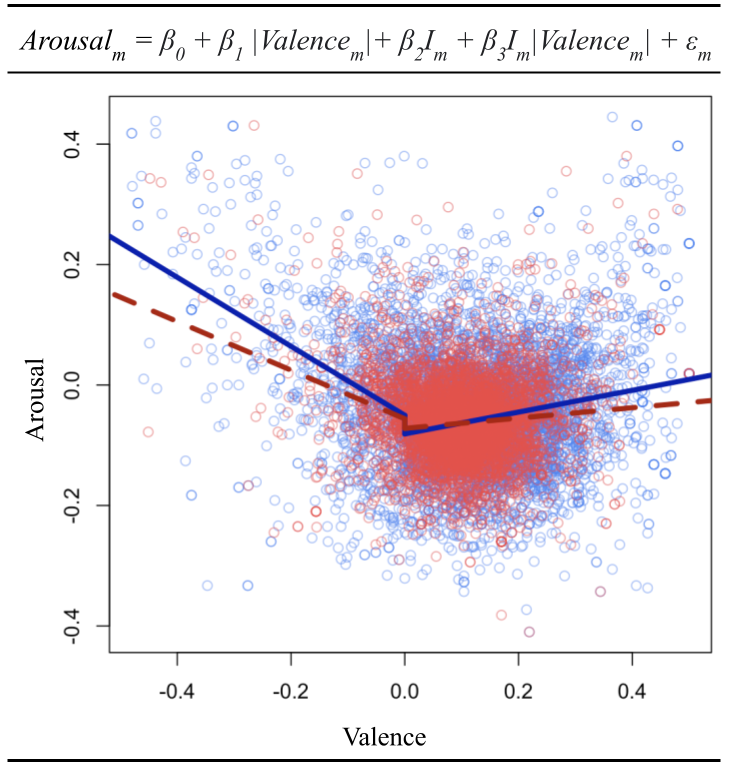}
    \caption{Scatter plots of valence (x-axis) and arousal (y-axis) of Twitter (blue) and Weibo (red) posts. The lines of best fit for Model 6's function are appended to the plot(Twitter: solid line, Weibo: dashed line). The model is tested with within-person intercept and slope.}
    \label{fig:model6}
\end{figure}

\begin{figure*}[!t]
    \centering
    \includegraphics[width=0.8\linewidth]{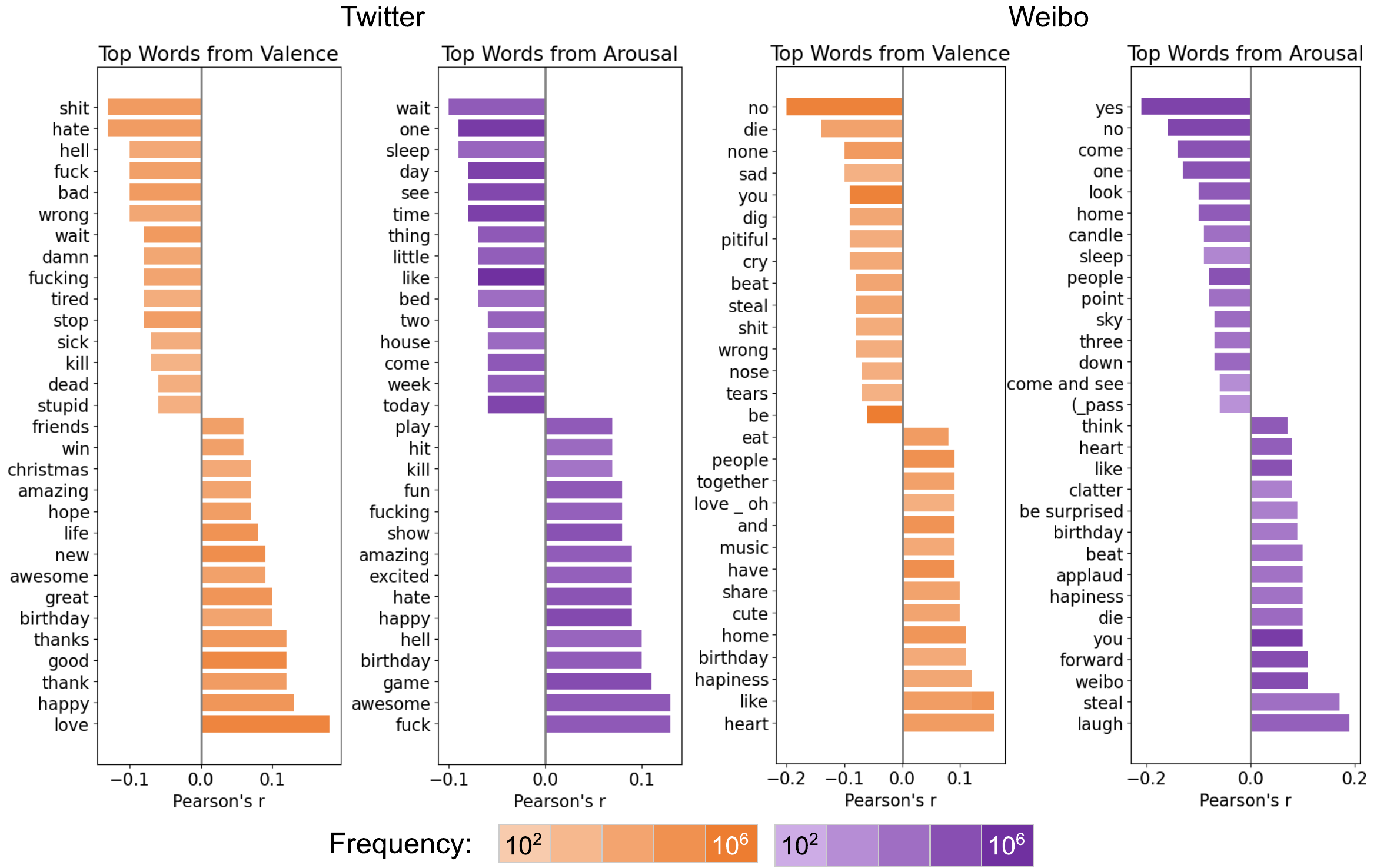}
    \caption{Words and phrases associated with valence and arousal on Twitter and Weibo (translated) from the top 15 phrases for effect strength (Pearson $r$), colored by frequency. Statistically significant ($p < .05$, two-tailed t-test, Benjamini-Hochberg corrected).}
    \label{fig:tw_wb_bar}
\end{figure*}

\section{Results}
\textcolor{red}{Warning: The following section contains swear words.}

\subsection{Valence-Arousal Functional Relationship}

The full comparison between 6 models are shown in Figure \ref{fig:func_rela_full}. Among the different models we tested across Twitter and Weibo data, Model 6 with within-person intercept and slope best fit with the lowest AIC and BIC, and highest Conditional $R^2$. Within-person models were also significantly different from the models without within-person effects.

\begin{table}[!b]
\centering
\resizebox{\columnwidth}{!}{%
\begin{tabular}{llcccc}
\hline
\textbf{Dataset}                  & \textbf{Model} & \textbf{AIC}      & \textbf{BIC} & \textbf{PostP} & $\mathbf{R^2}$\\ \hline
\multirow{6}{*}{\textbf{Twitter}} & \textbf{Model 1} & $-3.752 \times 10^6$ & $-3.752 \times 10^6$ & 0 & 0.015\\
& \textbf{Model 2} & $-3.790 \times 10^6$ & $-3.790 \times 10^6$ & 0 & 0.030\\
& \textbf{Model 3} & $-3.833 \times 10^6$ & $-3.833 \times 10^6$ & 0 & 0.050\\
& \textbf{Model 4} & $-4.001 \times 10^6$ & $-4.001 \times 10^6$ & 0 & 0.113\\
& \textbf{Model 5} & $-4.048 \times 10^6$ & $-4.048 \times 10^6$ & 0 & 0.135\\
& \textbf{Model 6} & $\mathbf{-4.060 \times 10^6}$ & $\mathbf{-4.060 \times 10^6}$ & \textbf{1} & \textbf{0.139}\\ \hline
\multirow{6}{*}{\textbf{Weibo}} & \textbf{Model 1} & $-4.155 \times 10^5$ & $-4.155 \times 10^5$ & 0 & 0.021\\
& \textbf{Model 2} & $-4.162 \times 10^5$ & $-4.161 \times 10^5$ & 0 & 0.025\\
& \textbf{Model 3} & $-4.172 \times 10^5$ & $-4.171 \times 10^5$ & 0 & 0.030\\
& \textbf{Model 4} & $-4.219 \times 10^5$ & $-4.219 \times 10^5$ & 0 & 0.057\\
& \textbf{Model 5} & $-4.247 \times 10^5$ & $-4.246 \times 10^5$ & 0 & 0.082\\
& \textbf{Model 6} & $\mathbf{-4.251 \times 10^5}$ & $\mathbf{-4.250 \times 10^5}$ & \textbf{1} & \textbf{0.084}\\ \hline
\end{tabular}%
}
\caption{Results of fitting 6 different models on Twitter and Weibo dataset. AIC is Akaike Information Criterion, BIC is Bayesian Information Criterion (the lower the better fit), PostP is posterior probability, $R^2$ is Conditional $R^2$.}
\label{tbl:func_rela_full}
\end{table}

As shown in Figure \ref{fig:model6}, the presence of an asymmetric V-shape in the data, including a negativity bias and negativity offset, was confirmed in the models on both Twitter and Weibo data. Compared with Weibo, Twitter shows a larger intercept gap (Twitter: $\beta_{2} = -0.031$; Weibo: $\beta_{2} = -0.016$). The intensity of emotion gets significantly stronger with higher positivity/negativity. This conclusion is consistent with both Twitter and Weibo, with the smallest BIC values in Model 6, characterized by a V shape (Twitter: $\beta_{1}= 0.573$, Weibo: $\beta_{1} = 0.404$) and negativity bias (Twitter: $\beta_{3} = -0.392$, Weibo: $\beta_{3} = -0.318$). The Twitter model has a steeper slope on both positive and negative valence compared to Weibo (Twitter: $\beta_{1} = 0.573$, $\beta_{3} = -0.392$; Weibo: $\beta_{1} = 0.404$, $\beta_{3} = -0.318$).

\subsection{Social Media Language Analysis}
To uncover the content differences in emotional expression across cultures, we utilized differential language analysis to obtain the most correlated n-grams and topics in each platform. Figure \ref{fig:tw_wb_bar} shows the top significantly correlated words and phrases with valence and arousal in both platforms. On the dimension of valence, Twitter users tended to use words conveying superlatives ('great', 'awesome', 'amazing') and festive celebrations ('birthday', 'Christmas', 'new', 'win') in expressing positive valence, while profanity ('shit', 'fuck'), negation ('hate', 'bad', 'wrong') and discomfort ('wait', 'tired', 'stop') were indicative of negative valence. Conversely, Weibo users commonly employed terms related to personal affect ('like', 'love', 'happiness') and emojis ('oh', 'heart') when expressing positive valence, whereas words indicative of negation ('no') and sorrow ('sad', 'cry') are prevalent in expressing negative valence. On the dimension of arousal, Twitter users expressed profanity ('shit', 'fuck') and interpersonal expressions ('awesome', 'amazing') for high arousal while using terms indicating low activities ('sleep', 'bed') and time-oriented description ('today', 'week', 'day', 'time') for low arousal. In contrast, Weibo users predominantly utilized positive emojis('steal-laugh', 'applaud') to convey high arousal, while employing affirmation ('yes'), negation ('no'), and sharing aspects of daily life ('home', 'sleep') to express low arousal. The version of Figure \ref{fig:tw_wb_bar} without using words in NRC-VAD is shown in the Appendix.

\begin{figure*}[t!]
    \centering
    \includegraphics[width=0.75\linewidth]{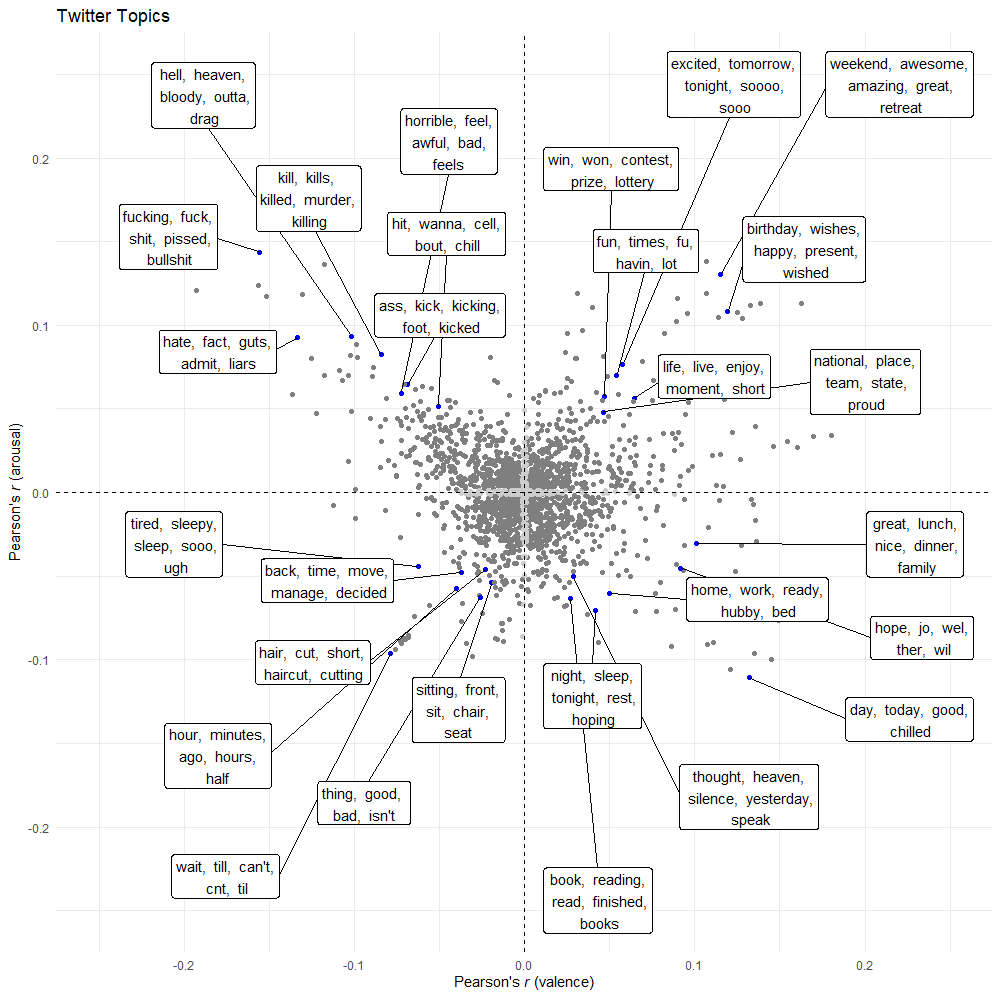}
    \caption{Topics associated with valence and arousal on Twitter, sorted by effect size (Pearson $r$).  Each point is a topic, and statistically significant topics ($p < .05$, two-tailed t-test, Benjamini-Hochberg corrected) are shown in dark gray. The X-axis is the Pearson $r$ with valence and the Y-axis with arousal. The top 5 words in each topic are shown. }
    \label{fig:tw_lda}
\end{figure*}

\begin{figure*}[t!]
    \centering
    \includegraphics[width=0.75\linewidth]{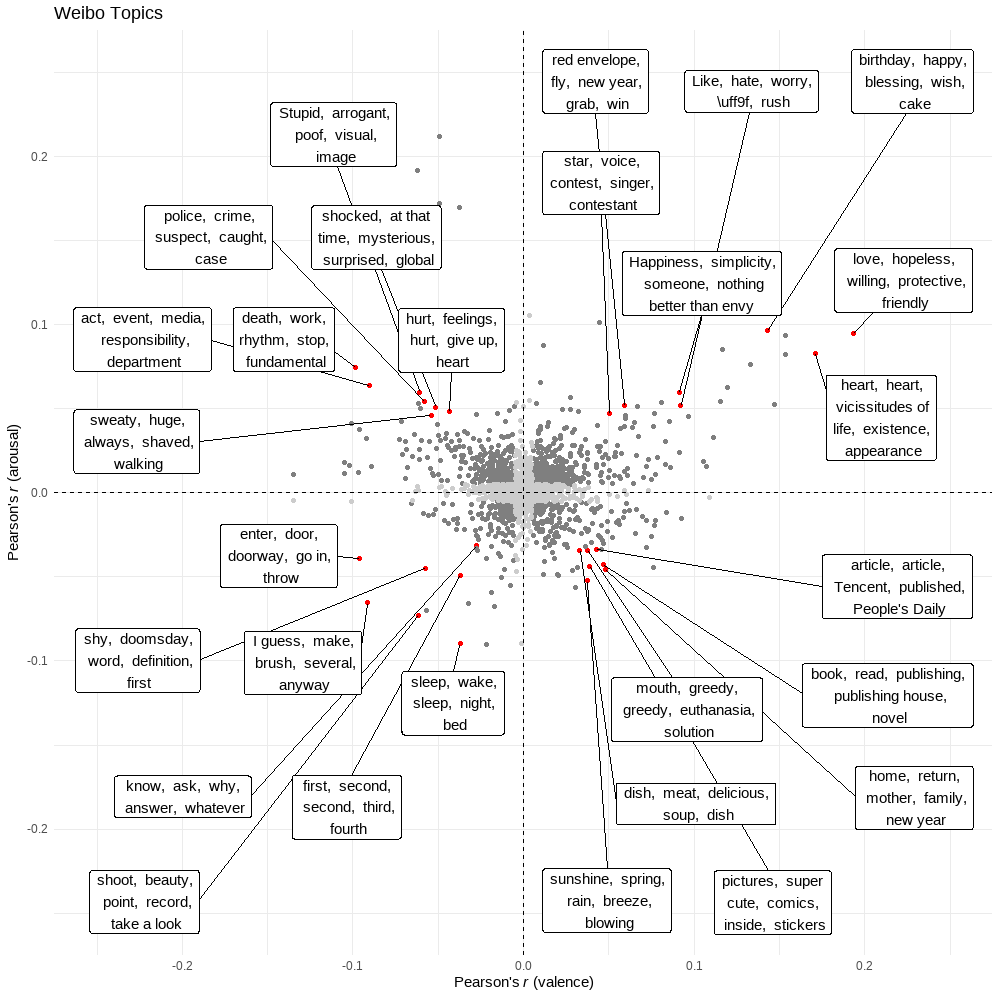}
    \caption{Topics associated with valence and arousal on Weibo, sorted by effect size (Pearson $r$).  Each point is a topic and statistically significant topics ($p < .05$, two-tailed t-test, Benjamini-Hochberg corrected) are shown in dark gray. The X-axis is the Pearson $r$ with valence and the Y-axis with arousal. English translations of the top 5 words in each topic are shown.}
    \label{fig:wb_lda}
\end{figure*}

We further compare Twitter and Weibo's LDA results for topics in Figure \ref{fig:tw_lda} and \ref{fig:wb_lda}. Twitter users had relaxing weekend ('weekend', 'awesome', 'amazing', 'great', 'retreat'), celebration of events ('birthday', 'wishes', 'happy', 'present', 'wished'), luck and achievement ('win', 'won', 'contest', 'prize', 'lottery') for positive valence high arousal. Conversely, Weibo users discussed affectionate bonding ('love', 'hopeless', 'willing', 'protective', 'friendly', where hopeless means love in deep) to express their feelings, particularly in the context of festivals and celebrations ('new year', 'red envelope') and interests in celebrities and TV shows ('celebrities','singer'). For positive valence low arousal, Twitter users usually talked about relaxing routines ('day', 'today', 'good', 'chilled') and sleep ('night', 'sleep', 'tonight', 'rest', 'hoping'). Besides, Weibo users shared family reunion ('home', 'return', 'mother', 'family', 'new year') and savory cuisines ('dish', 'meat', 'delicious', 'soup', 'dish'). When expressing strong negative feelings, Twitter users mainly used profanities ('fucking', 'fuck', 'shit', 'pissed', 'bullshit') to convey intense emotions, while Weibo users discussed law enforcement and criminal investigation ('police', 'crime', 'suspect', 'caught', 'case'). Additionally, Weibo discussions on negative high arousal included the use of emojis ('sweat') and negative emotions ('shocking', 'hurt', 'give up'). Concerning negative valence low arousal, Twitter users usually showed personal negative feelings like tiredness ('tired', 'sleepy', 'sleep', 'sooo', 'ugh'), engaged in discussions about daily activities ('hair', 'cut', 'short', 'haircut', 'cutting') and mentioned words related to time ('hour', 'minute'). Similarly, Weibo users also mentioned sleep ('sleep', 'awake', 'bed'). 

\section{Discussion}

This paper examined the functional relationship between valence and arousal based on large-scale social media texts across the United States and China. Our findings suggest that public affective expressions replicate the asymmetrical affective V-shaped relationship but with a negativity bias (negative feelings increase more strongly than positive feelings with increasing arousal) and negativity offset (feelings of arousal are higher at low negative valence levels than positive valence). In addition, the arousal and valence slope was steeper for Twitter users than for Weibo users.   

One of the major findings in our study is that the American participants had stronger negativity bias and overall had higher arousal with higher positive and negative valence compared to Chinese participants. This is consistent with past findings on West-East distinction in emotional arousal and aligns with Hofstede’s Individualism vs. Collectivism dimension: in Western or individualist culture, high-arousal emotions are valued and promoted more than low-arousal emotions, while in Eastern or collectivist culture, low-arousal emotions are valued more than high-arousal emotions \cite{lim2016cultural, hofstede2011dimensionalizing}. This can be attributed to the fact that individualistic culture encourages expressive independence and the externalization of personal emotions while collectivist culture values social harmony and group cohesion. Even in traditional Asian medicine, there is an assumption that excessive emotional expression can be harmful and cause diseases, whether it is positive or negative emotions \cite{lim2008emotional}. Our findings confirmed that Chinese users of Weibo express lower arousal levels for both negative and positive emotions. 

Content analyses of the findings suggested that Chinese participants displayed less high arousal positive affect emotional behavior than their American counterparts. This is consistent with past findings that there seems to be a general preference in the West for high-arousal positive states like excitement or enthusiasm \cite{sommers1984reported}. At the same time, people in the East generally prefer low-arousal positive affective states like calm or peacefulness \cite{tsai2007ideal}. Moreover, we saw Twitter users using more explicit excitement-focused terms such as awesomeness, while Weibo users tended to express positive emotions more implicitly, e.g., emojis. This is consistent with findings that the communication style of East Asian language communities tends to be more indirect than that of their Western counterparts \cite{fong1998chinese, gudykunst1988culture, neuliep2012relationship}.

Similarly, past literature suggests that high-arousal emotions serve as an effective means of influencing others in the West \cite{tsai2007ideal}, while low-arousal emotions serve as an effective means of adjusting and conforming to others in the East \cite{markus1991cultural}. We found that low arousal emotions in Weibo were used to create a sense of comfort and connection through themes related to nature, family, daily life activities, and light-hearted entertainment. On the high arousal-high positive affect sphere, Twitter users celebrated more personal events, while Weibo users talked more excitedly about celebrities and current events. Therefore, it is likely that while the ideal affect preference translates into affective expressions about personal experiences in the East, discussion of media culture is exempt from such norms: for instance, while it may be frowned upon to act too excited about personal events, the same restrictions are not in place when expressing excitement about celebrities and cultural events. As such, our findings provide a novel insight into our understanding of norm differences in affective expression in East vs West. 

Similarly, looking at the difference in negativity bias for Twitter and Weibo, while Twitter users use profanity primarily, Weibo users tend to use words with much lower intensity, confirming the assumption that Chinese users try to avoid expressing extreme emotions.
Note that although Weibo has censorships, it does not include profanity filters.

One surprising finding in our study was that we did not find a positivity offset. We instead found a negativity offset for both American and Chinese participants. The theoretical explanation for the positivity offset (and negativity bias) comes from the Evaluative Space Model (ESM; \citealp{cacioppo1999affect, norris2010current}), which proposed that positive and negative affect have different arousal functions and predicts greater positive than negative affect at low levels of affective input. The adaptive reason for the offset was hypothesized to encourage approaching novel stimuli in low-threat conditions. However, our finding suggests this may not translate to public affective behavior, particularly on social media. It suggests that people on both Twitter and Weibo are more likely to approach neutral stimuli in negative terms while simultaneously having stronger negative reactions to higher arousal events. Therefore, our studies elucidate how certain theories of affect may not explain affective behavior universally, partly because of the contexts not considered in said theories. 
\section{Conclusion}
This study highlights the importance of studying public emotional behavior and how it is distinguished from self-reported findings. Our findings could confirm some theoretical assumptions in traditional self-report research by adding new empirical evidence when applied to public emotional behavior. Future research looking at individual self-reports and public behavior can help us understand what these differences can represent at the individual level.

\section{Limitations}
\textbf{Platform Issue:} Even though Twitter and Weibo are comparable in usage \cite{li2020studying,guntuku2019studying} and have not been shown to have significant differences in predicting individual states \cite{gao2012comparative}, data from other platforms such as WeChat and RenRen in China and Facebook in the US have not been included in this study due to access constraints. Emojis are a significant contributor to affective expressions \cite{li2019exploring}; however, we did not include them in this study due to differences in encodings while collecting the data making it infeasible for us to parse them accurately. Further, social media users are non-representative of the general population, and the participants in this study are non-random and convenient samples. 

\textbf{Fine-grained Emotion:} Our focus in this paper was to compare the expressions of valence and arousal across two different cultures building upon rich cross-cultural psychological studying the difference in valence and arousal \cite{lim2016cultural, kuppens2017relation}. Although considering fine-grained emotions could make the analysis multidimensional, it will make the results less reliable. Moreover, each of the fine-grained emotions could be represented in the valence and arousal circumplex \cite{jefferies2008emotional, mohammad-2018-word}.

\textbf{Subcultural Variance:} We acknowledge the existence of subcultural variances, such as those among various ethnicities, provinces, and counties in China and the US. For example, minority students, including Tibetan and Mongolian, tend to experience more negative emotions and are less inclined to adopt emotion regulation strategies compared to Han students \cite{lu2012emotional}. Within the United States, European Americans show a greater motivation to engage in hedonic emotion regulation than their Asian American counterparts \cite{miyamoto2014cultural}. 

However, despite these nuances, the macro-cultural differences between East and West remain significant enough to warrant a comparative analysis \cite{lu2001cultural, tsai2006cultural}. Focusing on broader cultural variation, our investigation emphasizes the pronounced disparities between the US and Chinese cultural contexts by using social media posts. These disparities are substantial and provide a robust framework for comparative analysis. This macro-level perspective is not to negate the relevance of subcultural variances but to highlight the overarching patterns that emerge when contrasting Eastern and Western cultures by using two large countries that have a large variation in Hofstede’s cultural dimensions, for instance, as examples. By situating our work within this broader context, we aim to contribute to a more comprehensive understanding of how culture influences emotional expression. In the discussion, we will add the above language to acknowledge the complex variety of cultural diversity that exists within and across national borders.

\textbf{Translation:} Lexica need to be adapted to the cultures to measure psychological phenomenon accurately. We tried using Chinese valence-arousal words (CVAW, \citealp{lee2022chinese}). However, we did not proceed further as the methods of building NRC-VAD (for English) and CVAW (for Chinese) lexica were different and could cause misalignment. We wanted to control for such differences by choosing a lexicon that has sufficient coverage while also being used in multiple prior works across both languages \cite{wenjia-etal-2023-improving, mohammad2016sentiment, li2019exploring, chen2019sentimental, das2021characterizing}. Further, with over 20K entries, NRC-VAD is the largest manually created emotion lexicon that has translations across several languages. 

\textbf{Censorship:} We acknowledge that censorship on Sina Weibo is a challenge. Despite this issue, Weibo has been successfully used across multiple studies to understand different psychological outcomes (e.g. affect, stress, depression; \citealp{pan2021china, tang2019effects, li2014predicting}) 

\section{Ethics}
The University of Pennsylvania’s Institutional Review Board declared this project exempt (IRB protocol \# 829811).

This study, focusing on the cultural differences in affective expressions between Twitter users in the United States and Sina Weibo users in China, raises several ethical considerations:  

\textbf{1. Data Privacy and Anonymity}: The research analyzes social media posts from Twitter and Sina Weibo. It is important to ensure that individual users' privacy is respected. All data extracted from these platforms is anonymized by removing personally identifiable information.  

\textbf{2. Cultural Sensitivity and Bias}: Given the cross-cultural nature of the study, it is critical to approach the analysis with cultural sensitivity. Researchers must be aware of and mitigate any biases arising from their cultural backgrounds or perspectives. This includes being mindful of how cultural contexts influence affective expressions and the interpretation thereof.  

\textbf{3. Representation and Generalization}: Care should be taken to avoid over-generalizing the findings. The study's results are based on specific social media platforms and may not represent the broader United States and China populations.

\bibliography{custom}

\section*{Appendix}
\appendix
\label{sec:appendix}

\section{Details of Pearson \texorpdfstring{$r$}{r}}
\label{app:pearson}

In our paper, Pearson $r$ correlation is independently calculated for post level valence and arousal scores, which gives each n-gram and topic a valence and arousal score. The Pearson r correlation coefficient is calculated with the OLS regression. Since the Coefficient in OLS regression is:
\begin{equation}
    \beta = \frac{Cov(x,y)}{Var(x)}
\end{equation}
And Pearson Correlation Coefficient is:
\begin{equation}
    r = \frac{Cov(x,y)}{\sqrt{Var(x)\cdot Var(y)}}
\end{equation}
So $\beta$ can be represented by $r$ with:
\begin{equation}
    \beta = r \cdot \frac{SD(y)}{SD(x)}
\end{equation}
So when normalized, $\beta = r$.
\begin{figure}[!t]
    \centering
    \includegraphics[width=\linewidth]{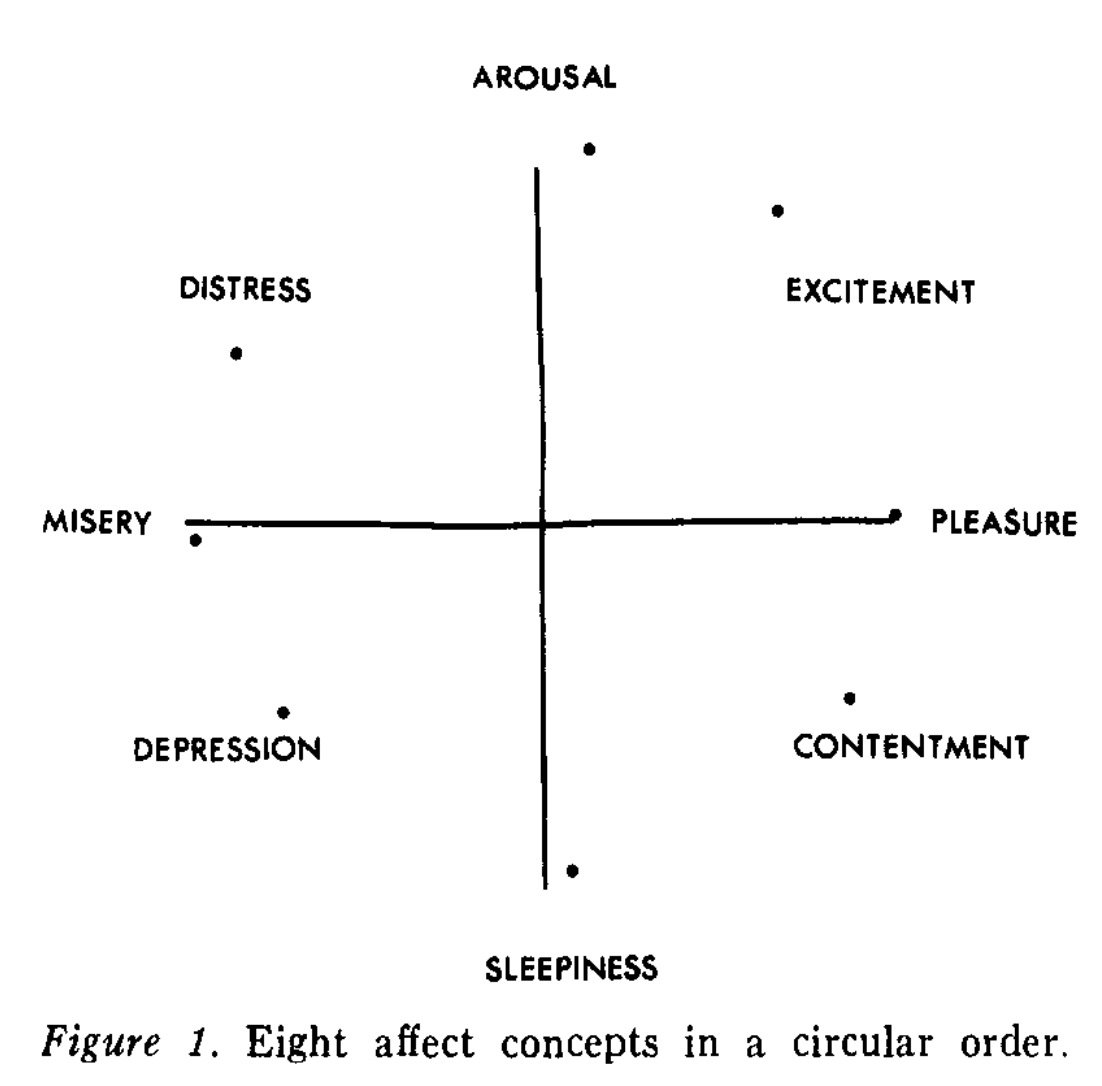}
    \caption{Example of Russell's two-dimensional circumplex model. Captured from \citealt{Russell1980-au}.}
    \label{fig:russell}
\end{figure}
\section{Explanation of Conditional \texorpdfstring{$R^2$}{R2}}
In Table \ref{tbl:func_rela_full}, we use conditional $R^2$ to represent the variance explained by the entire model, including both fixed and random factors.

\begin{figure}[!t]
    \centering
    \includegraphics[width=\linewidth]{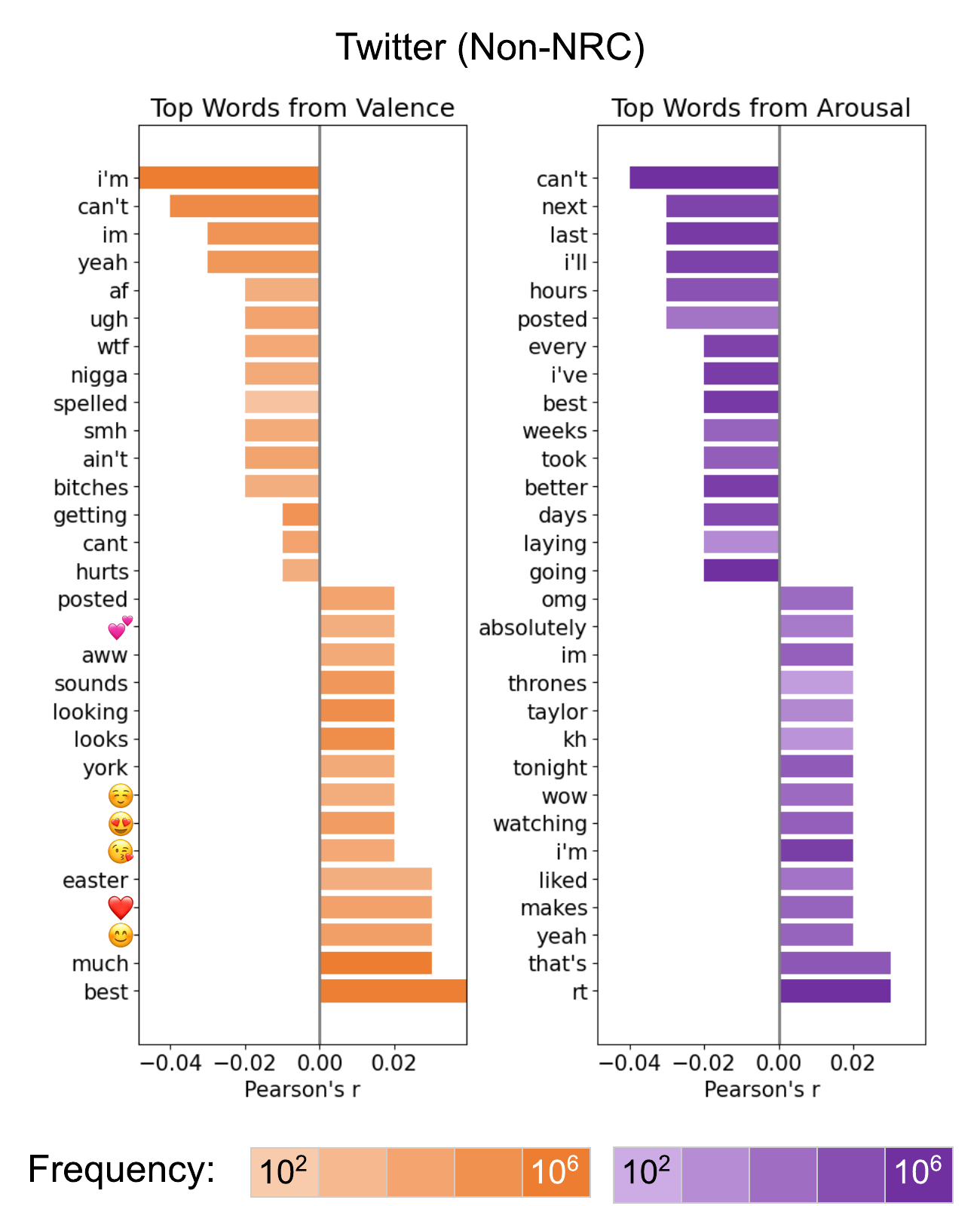}
    \caption{Words and phrases associated with valence and arousal on Twitter (Non-NRC) from the top 15 phrases for effect strength (Pearson $r$), colored by frequency. Statistically significant ($p < .05$, two-tailed t-test, Benjamini-Hochberg corrected).}
    \label{fig:tw_no_NRC}
\end{figure}

The method we used for finding functional relationships between valence and arousal is a follow-up analysis using the same method used in the previous work. Table 2 in \citealp{Kuppens2013-im} listed BIC, PostP, and the best $R^2$ for different studies and datasets, which are consistent with our result - relatively low $R^2$ scores. In our paper, we expand this work using social media data to see if a similar conclusion can hold across cultures with the functional relationship between valence and arousal.

As mentioned in \cite*{Kuppens2013-im}, low to moderate $R^2$ values of our functions are expected, not only because our analysis is based on noisy social media data, but also affective experiences of all combinations of valence and arousal can occur. For example, although less likely, the LDA result in our paper shows that positive valence low arousal states are represented by relaxing, and sleep.

Our goal in the analysis of functional relationships is not to train a model for predicting the arousal of a sentence using valence, but to explain the relationship between valence and arousal on average. 

\begin{figure}[!t]
    \centering
    \includegraphics[width=\linewidth]{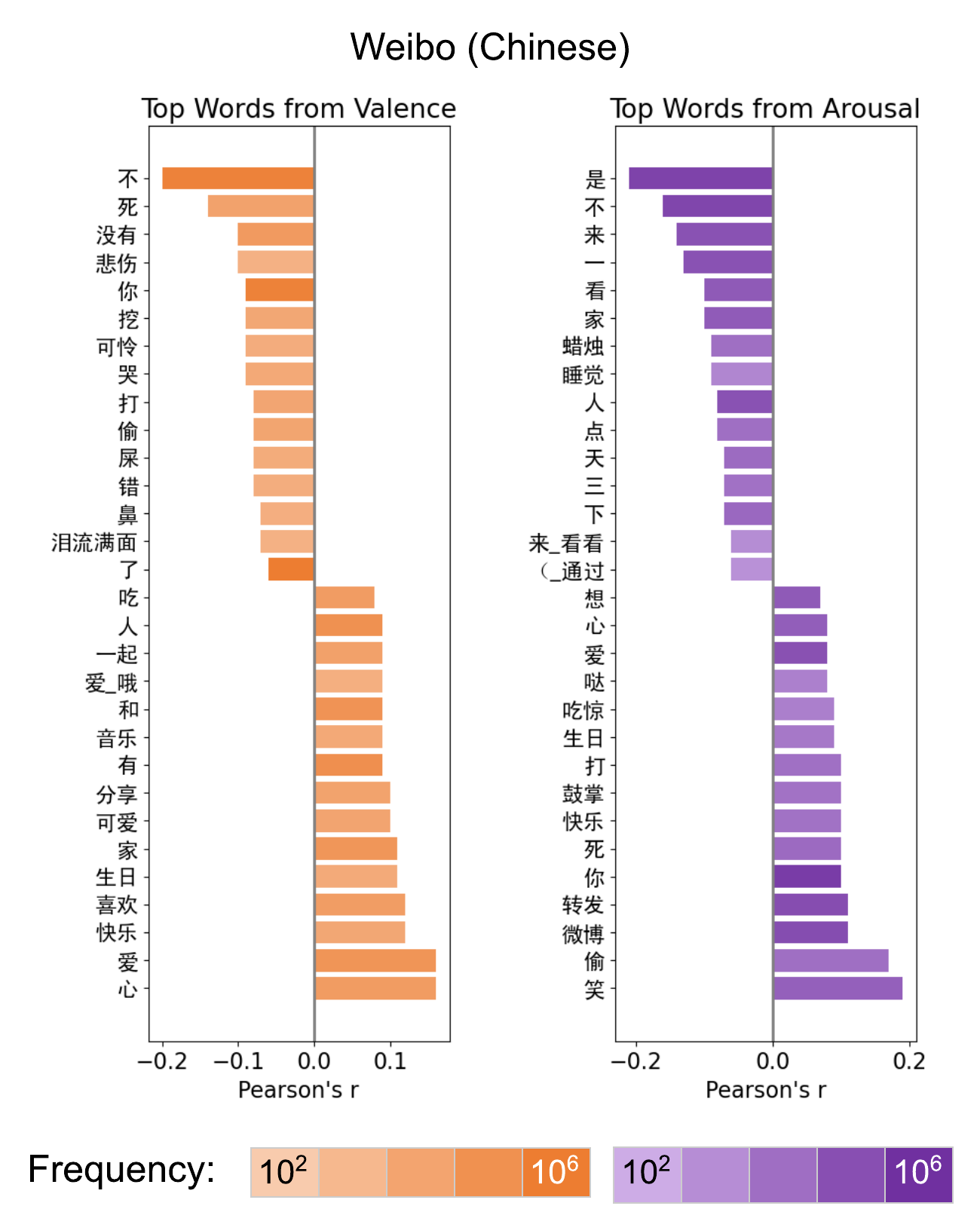}
    \caption{Words and phrases associated with valence and arousal on Weibo (Chinese) from the top 15 phrases for effect strength (Pearson $r$), colored by frequency. Statistically significant ($p < .05$, two-tailed t-test, Benjamini-Hochberg corrected).}
    \label{fig:wb_zh_bar}
\end{figure}

\begin{figure*}[t!]
    \centering
    \includegraphics[width=0.75\linewidth]{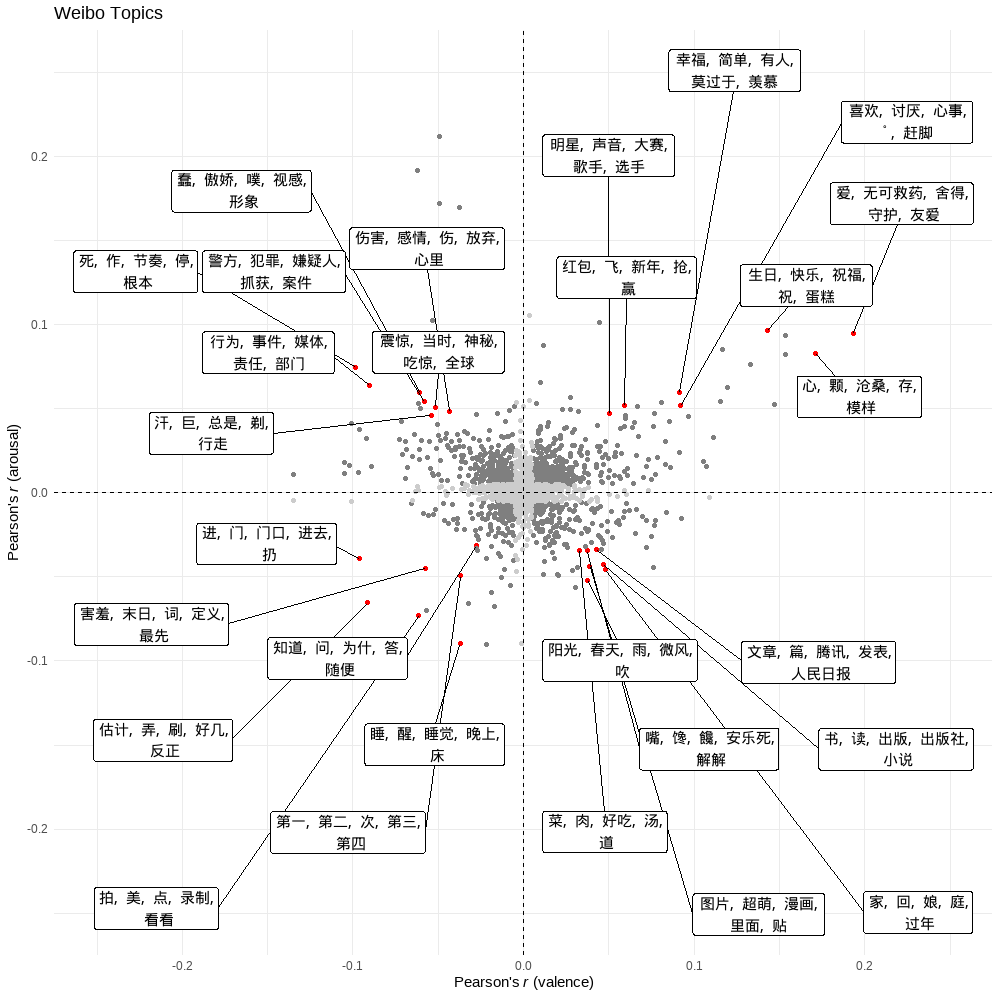}
    \caption{Topics associated with valence and arousal on Weibo (Chinese), sorted by effect size (Pearson $r$).  Each point is a topic and statistically significant topics ($p < .05$, two-tailed t-test, Benjamini-Hochberg corrected) are shown in dark gray. The X-axis is the Pearson r with valence and the Y-axis with arousal. English translations of the top 5 words in each topic are shown.}
    \label{fig:wb_zh_lda}
\end{figure*}

\section{Details of Valence-Arousal Relation Models}
\begin{figure*}[t]
    \centering
    \includegraphics[width=\linewidth]{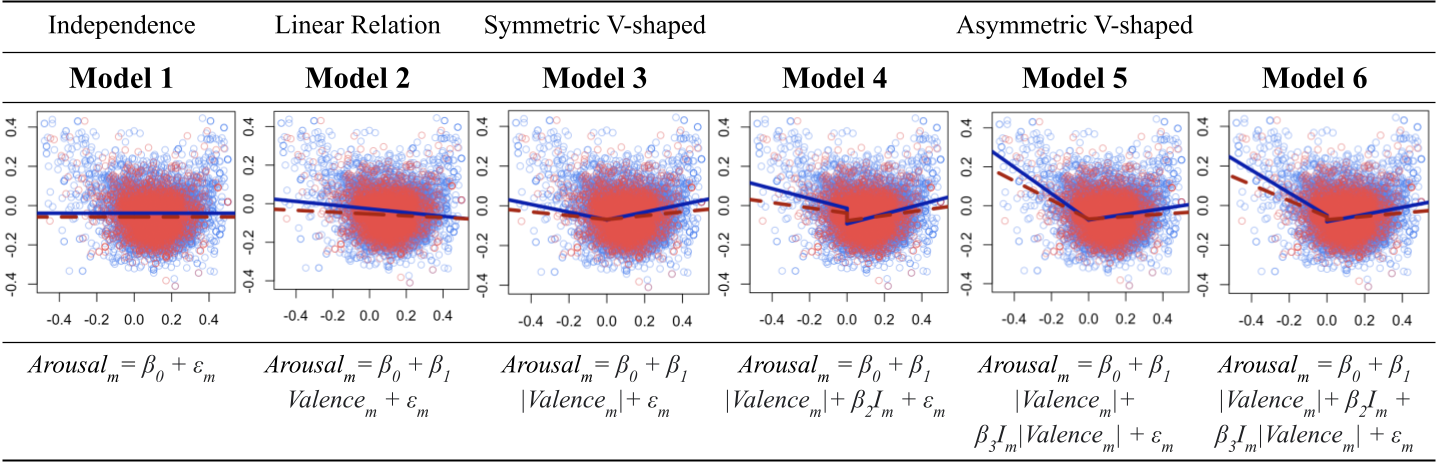}
    \caption{Scatter plots of valence (x-axis) and arousal (y-axis) of Twitter (blue) and Weibo (red) posts. The lines of best fit for each model's function are appended to each plot (Twitter: solid line, Weibo: dashed line). Each model is tested with a within-person intercept and slope.}
    \label{fig:func_rela_full}
\end{figure*}

In this section, we show the full comparison between Model 1 to Model 6 in Figure \ref{fig:func_rela_full}. Among all, Model 6 fits the social media data best, shows negative offset and negative bias in the line of best fit. This is opposite to the previous findings where positive offsets were observed from lab-based measurements. 

\section{Removal of NRC-VAD Lexicons}
To give a multidimensional insight into culturally specific expressions of valence and arousal, we show Twitter part of Figure \ref{fig:tw_wb_bar} without using words from NRC-VAD Lexicon in Figure \ref{fig:tw_no_NRC}. NRC-VAD contains 20,000 words, which covers most of the daily vocabulary. This figure, without lexicon words, consists of a lot of emojis, internet slang, and swear words. 

\section{Pre-Translated Weibo Figures}
All the figures and tables from Weibo are translated into English with Google Translate. Here, we show the figures with original Chinese text. Figure \ref{fig:wb_zh_bar} shows the top 15 phrases for effect strength with valence and arousal. Figure \ref{fig:wb_zh_lda} shows the top topics associated with valence and arousal.

\end{document}